\documentclass[twocolumn, preprintnumbers, eqsecnum, letterpaper, superscriptaddress, nofootinbib]{revtex4-1}

\usepackage{graphicx}
\usepackage{microtype}
\usepackage{amsmath}
\usepackage{amssymb}
\usepackage{subfigure}
\usepackage{hyperref}
\usepackage{url}
\usepackage{xcolor}
\usepackage{color}
\usepackage{mathrsfs}
\usepackage{calrsfs}
\usepackage{amsfonts}
\usepackage{eufrak}
\usepackage{tabularx}
\usepackage{eucal}
\usepackage{latexsym}
\usepackage{ragged2e}
\usepackage{epsfig}
\usepackage{textcomp}
\usepackage{caption}
\usepackage{subfigure}
\usepackage{marvosym}
\usepackage{ifsym}
\usepackage{lipsum}

\DeclareCaptionJustification{justified}{\leftskip=0pt \rightskip=0pt \parfillskip=0pt plus 1fil}
\definecolor{vividviolet}{rgb}{0.62, 0.0, 1.0}
\definecolor{amaranth}{rgb}{0.9, 0.17, 0.31}
\definecolor{palatinateblue}{rgb}{0.15, 0.23, 0.89}
\definecolor{brightpink}{rgb}{1.0, 0.0, 0.5}
\definecolor{cornflowerblue}{rgb}{0.39, 0.58, 0.93}
\definecolor{deepcarminepink}{rgb}{0.94, 0.19, 0.22}
\definecolor{radicalred}{rgb}{1.0, 0.21, 0.37}

\hypersetup{linktoc=all,
    colorlinks, linkcolor={palatinateblue},
    citecolor={brightpink}, urlcolor={amaranth}
}

\graphicspath{{Images/}}

\graphicspath{{Images/}}

\def\@fnsymbol#1{\ensuremath{\ifcase#1\or \ddagger \or  $\textleaf$  \or \dagger
\else\@ctrerr\fi}}%

\makeatother

\def\sideremark#1{\ifvmode\leavevmode\fi\vadjust{\vbox to0pt{\vss
 \hbox to 0pt{\hskip\hsize\hskip1em
 \vbox{\hsize1.3cm\tiny\raggedright\pretolerance10000
 \noindent #1\hfill}\hss}\vbox to8pt{\vfil}\vss}}}%

\def\beq{\begin{equation}}
\def\eeq{\end{equation}}

 \newcommand{\be}{\begin{equation}}
	\newcommand{\en}{\end{equation}}

\begin{document}

\title{The FRW Universe as a van der Waals-like Thermodynamic Heat Engine}



\author{Haximjan Abdusattar \Letter}
\email{axim@ksu.edu.cn}
\affiliation{School of Physics and Electrical Engineering, Kashi University, Kashi 844000, Xinjiang, China}

\author{Sa\"{u}l Tom\`{a}s V\'{a}zquez}
\email{saulrialp@gmail.com}
\affiliation{Theoretical Researcher,  {\AE}ther Project, Rialp, 25594, Spain}

{\let\thefootnote\relax\footnotetext{\vspace*{0.1cm}$^{\text{\Letter}}$ Corresponding Author}}


\begin{abstract}

It is well known that the Friedmann-Robertson-Walker (FRW) universe is a dynamical spacetime, and it has thermodynamics embodied on the apparent horizon. Notably, it also possesses a van der Waals-like equation of state, enabling us to consider thermodynamic cycles and explore the potential of the FRW universe as a heat engine. In this paper, we investigate two types of cycle--the Carnot cycle and the rectangular cycle--based on the phase diagram derived from the equation of state, to study the heat engine characteristics of the FRW universe. Furthermore, we calculate the work done and assess the corresponding efficiencies, illustrating the efficiency diagram for the FRW universe's heat engine. We observe that the efficiency of the rectangular cycle consistently remains below unity and never exceeds the Carnot efficiency--the thermodynamic upper limit. This finding is in alignment with the traditional thermodynamic principles that govern heat engines.

\end{abstract}

\maketitle

\section{Introduction}

Black hole thermodynamics is a fruitful area in modern theoretical physics, which has inspired several cutting edge studies. Black hole phase transition and critical phenomenon takes a pivotal status in black hole thermodynamics. In explorations of critical phenomenon and phase transition of black holes, the thermodynamic pressure plays a key role. In recent years, the cosmological constant $\Lambda$ is treated as
the thermodynamic variable analogous to the pressure $P_\Lambda=-{\Lambda}/{8\pi}$ \cite{Kastor:2009wy,Dolan:2010ha} and its conjugate quantity is the thermodynamic volume, which yields the presence of a pressure-volume term in the first law of thermodynamics. In this framework, the idea of pressure and volume as well as the thermodynamic equation of state in the extended phase space of black hole thermodynamics in Anti-de Sitter (AdS) spacetime has drawn a lot of attention
\cite{Kubiznak:2012wp,Kubiznak:2016qmn,Wei:2012ui,Gunasekaran:2012dq,Hendi:2012um,Cai:2013qga,Altamirano:2014tva,Dehghani:2014caa,Xu:2015rfa,Cheng:2016bpx,Hu:2018qsy,Estrada:2019cig,Li:2020xkh,Abdusattar:2023xxs,Abdusattar:2024zzi,Abdusattar:2025rdp}.

Both thermodynamics and gravity exhibit an extraordinary universality. The universe, as the largest known system governed by gravity, ought to adhere to the laws of thermodynamics. In contrast to typical cases of black hole thermodynamics, the Friedmann-Robertson-Walker (FRW) universe represents a dynamical system. Some studies indicate that the FRW universe has thermodynamics embodied on the apparent horizon, whose area plays the role of the entropy, and whose surface gravity behaves as temperature. In spirit of Jacobson's derivation \cite{Jacobson:1995ab} of the Einstein field equations from the Clausius relation, Cai and Kim \cite{Cai:2005ra} first investigated the Friedmann equations of the FRW universe on the apparent horizon, whose thermodynamics is associated with the unified first law \cite{Cai:2006rs}. A similar connection between the Friedmann equations and the first law of thermodynamics has also been discovered in alternative theories of gravity \cite{Akbar:2006kj,Akbar:2006mq}.

The other significant property of thermodynamics of FRW universe is that it deal with a non asymptotically flat spacetime. The total values of some extensive quantities like mass and entropy generally do not make sense in an FRW universe, except the universe with positive spatial curvature. It inevitability fails if one just naively mimics the black hole thermodynamics to investigate the thermodynamics of FRW universe, since the extensive thermodynamic quantities such as mass and entropy are taken as the total ones of the manifold in black thermodynamics. Thus the thermodynamics of FRW universe requires the values of extensive quantities in a finite region. We are led to the notorious problem, the delocalization of gravitational energy. A quasi local energy of gravity field must be invoked \cite{Hayward:1994bu,Hu:2015xva}. There are several different definitions of quasi local energies of gravity field. For an asymptotically flat manifold, they converge to the Arnowitt-Deser-Misner (ADM) mass at spacelike infinity. But for a non asymptotically flat manifold, they are different on the total manifold. Thermodynamical laws provide an interesting approach to probe which one is more reasonable in such a dynamical spacetime \cite{Hayward:1993wb,Hayward:1997jp}. Moreover, research into the thermodynamics of the FRW universe provides insights into the inherent relationship between thermodynamics and gravity, offering a universal and profound understanding of the nature of gravity.

However, to our knowledge, the thermodynamics of the FRW universe has yet to be systematically explored, primarily due to the lack of clarity regarding an effective gravitational pressure. Nonetheless, initial attempts have been made to study this topic by drawing parallels with the thermodynamics of black holes. In \cite{Debnath:2020inx}, the author study the thermodynamic properties of the FRW universe, in which the universe has a constant apparent horizon, or $\dot{R}_A\approx 0$.\footnote{In \cite{Bhadra:2024xcp}, the assumption $\dot{R}_A = R_A H$, despite its mathematical convenience, imposes a stringent constraint on the apparent horizon's variation. Specifically, it leads to $\dot{H}=0$ (derived from Eq.(5)) and consequently, through Eq.(21), results in $\rho+3p=0$.}
Generally, this condition is not satisfied in the realistic universe. Due to dynamical nature of the FRW universe, we
should consider time-dependence of the apparent horizon, which is also related to the Hawking temperature.
In our recent works \cite{Abdusattar:2021wfv,Kong:2021dqd,Abdusattar:2022bpg}, we have investigated the varying apparent horizon and extracted the definition of the thermodynamic pressure from the first law of thermodynamics. With such definition, we have further obtained the thermodynamic equation of state for a FRW universe. Once the thermodynamic equation of state is obtained, we can discuss its thermodynamic properties as the universe evolves \cite{Abdusattar:2024alq,Abdusattar:2024sgk,Abdusattar:2023pck}.

The presence of a pressure term allows us to consider thermodynamic cycles and the behavior of FRW universe as a heat engine. In this respect, it is possible to extract the mechanical work from the heat
energy via the term $P dV$. This suggests that the concept of the traditional heat engine can be
incorporated into the FRW universe at which the FRW universe play the role as the working substances. Heat engine was first idealized by Johnson \cite{Johnson:2014yja} and has been explored in different contexts \cite{Johnson:2015fva,Chandrasekhar:2016lbd,Liu:2017baz,Mo:2018hav,Hendi:2017bys,EslamPanah:2018ums,Ghaffarnejad:2018gtj}. More recently, Shahjalal et al.\cite {Shahjalal:2019pqb} conducted a systematic investigation of the thermodynamics of a quantum-corrected black hole embedded in a perfect fluid, with the latter acting as the source of the pressure term. In \cite{Bezerra:2019qkx}, the authors studied quantum effects on the criticality and efficiency of black holes surrounded by a perfect fluid. FRW universe as a thermodynamic system can be regarded as perfect fluid, in which we treat the work density of the perfect fluid as generator of the thermodynamic pressure term. Therefore, it would be interesting, and non-trivial to find the corresponding results in thermodynamics of the FRW universe. Motivated by these works, we study the properties of FRW universe as a thermodynamic heat engine and discuss its efficiencies as the universe evolves.

At this point, it is important to emphasize that the thermodynamic interpretation adopted throughout this work should be understood as an effective description. In a globally dynamical spacetime such as the FRW universe, notions like temperature, pressure, work, and thermodynamic cycles are expected to remain physically meaningful only within regimes where the apparent horizon evolves sufficiently slowly and quasi-equilibrium conditions can be assumed. Outside such regimes, thermodynamic quantities may still be formally defined, but their physical interpretation becomes conditional rather than global. The heat-engine picture explored in this paper should therefore be interpreted within this domain of applicability.

The organization of this paper is as follows. In Sec.\ref{sII}, we review thermodynamics of FRW universe on the apparent horizon. In Sec.\ref{sIII}, we present the efficiency of the heat engine for FRW universe as application of the thermodynamic equation of state. In
Sec.\ref{sIV}, we present conclusions and discussion of this paper.
Throughout this paper, we adopt natural units such that $c=G=\hbar=k_{B}=1$.

\section{Warm-Up: A Review of Previous Results on FRW Universe Thermodynamics}\label{sII}

As we embark on the intricate and fascinating journey of exploring the FRW universe as a heat engine, it is of paramount importance to gain a profound understanding of the underlying thermodynamic principles and the equation of state. This section aims to lay the cornerstone for this exploration by briefly reviewing the thermodynamics of the FRW universe and its unique equation of state, with a particular focus on the intrinsic relationship between thermodynamic pressure and work density. Here, let us together unveil a glimpse of the thermodynamics of the FRW universe, paving a solid theoretical foundation for subsequent in-depth studies on its characteristics as a heat engine.

\subsection{apparent horizon and Hawking temperature}

In the following, we will make a review of the apparent horizon, the Hawking temperature of FRW universe. In an isotropic coordinate system $x^\mu=(t, r, \theta, \varphi)$, the line element of FRW universe given by
\begin{equation}\label{FRW}
ds^2=-dt^2 +a^2(t)\Big[\frac{dr^2}{1-kr^2}+r^2 (d\theta^2 + \sin^2 \theta d\varphi^2)\Big]\,,
\end{equation}
where $a(t)$ is scale factor describing the evolution of the universe, $k=+1, 0, -1$ are the spatial curvatures corresponding to a spherical, flat and hyperbolic universe, respectively. If we introduce the areal radius $R(t,r)\equiv a(t)r$, the metric (\ref{FRW}) can be rewritten as:
\begin{equation} \label{nFRW}
ds^2=h_{ij}dx^{i} dx^{j}+R^2(t,r)(d\theta^2 + \sin^2 \theta d\varphi^2),
\end{equation}
where $x^{0}=t$, $x^{1}=r$, $h_{ij}={\rm diag}\big[-1,~\frac{a^2(t)}{1-kr^{2}}\big]$, and in what follows we may simply denote $a=a(t)$, $R=R(t,r)$. The FRW universe is a dynamical spacetime with an apparent horizon, which is a trapped surface with vanishing expansion $h^{i j}\partial_{i}R\partial_{j}R=0$.
From this relation, the apparent horizon $R_{A}$  can be solved as \cite{Cai:2005ra}
\begin{equation}\label{AH}
R_{A}=\frac{1}{\sqrt{H^{2}+\frac{k}{a^{2}}}} \,,
\end{equation}
where $H=H(t):=\dot a/a$ is the Hubble parameter characterizing the expansion rate of the universe. Taking the derivative of $R_{A}$ with respect to the cosmic time $t$, we obtain
\begin{equation}\label{dAH}
\dot{R}_{A}=-R_{A}^{3}H\Big(\dot{H}-\frac{k}{a^{2}}\Big) \,,
\end{equation}
which describes the nature of time dependence of the apparent horizon.

For dynamical spacetime, the surface gravity on the apparent horizon is defined by $\kappa={\partial _i}(
\sqrt{-h}~h^{ij}{\partial _{j}R})/(2\sqrt{-h})$ where $h={\rm det}(h_{ij})$ \cite{Cai:2006rs,Akbar:2006kj}. After simple calculation, one can obtain the surface gravity of FRW universe on the apparent horizon is $\kappa=-(1-{\dot{R}_{A}}/{2H R_{A}}
)/R_{A}$. Furthermore, the Hawking temperature of the FRW universe on the apparent horizon is \cite{Cai:2008gw}
\begin{equation}\label{T1}
T\equiv\frac{|\kappa|}{2\pi}=\frac{1}{2\pi {R_A}}\Big(1-\frac{\dot{R}_A}{2H R_{A}}
\Big) \,.
\end{equation}

\subsection{Thermodynamic Equation of State For the FRW Universe}

In the following, we discuss the thermodynamic equation of state of the FRW universe in Einstein gravity.
To construct the equation of state, we should find a proper definition of the thermodynamic
pressure $P$ and its conjugates. Therefore, we will start from the first law of thermodynamics of the FRW universe, in which the Misner-Sharp energy plays an important role. The Einstein field equations in the presence of cosmological constant can be written as
\begin{equation}\label{Eeq}
\mathcal {R}_{\mu\nu}-\frac{1}{2}g_{\mu\nu}\mathcal {R} =8\pi T_{\mu \nu}=8\pi(T_{\mu \nu}^m+T_{\mu \nu}^{\Lambda})\,,
\end{equation}
where $\mathcal {R}_{\mu\nu}$ is the Ricci tensor, $\mathcal {R}$ is the Ricci scalar, $T_{\mu \nu}$ is the total energy momentum tensor of the matter field inside the FRW universe which consisting of two parts:
the effective energy-momentum of the cosmological constant $T_{\mu \nu}^{\Lambda}=-\frac{\Lambda}{8\pi} g_{\mu \nu}$ and the energy-momentum tensor of the perfect fluid
\begin{equation}\label{Tmu}
T_{\mu \nu}^m=(\rho_m+p_m)u_{\mu}u_{\nu}+p_m g_{\mu \nu}\,,
\end{equation}
where $u^\mu$ denotes the four-velocity of the fluid and is given by $u^\mu=\frac{1}{\sqrt{-g_{tt}}}\delta_t^\mu$, and $\rho_m$, $p_m$ are the energy density and pressure, respectively.
For the FRW metric (\ref{nFRW}) and energy-momentum tensor (\ref{Tmu}), the first two components of the Einstein field equation can be written as~\cite{Ha:2009ft}
\begin{eqnarray}
\tilde{\rho}&=&\rho_{m}+\rho_{\Lambda}= \frac{3}{8 \pi}\Big(H^2+\frac{k}{a^2} \Big), \label{FE1} \\
\tilde{P}&=&p_{m}+P_{\Lambda} \nonumber\\
&=&-\frac{3}{8 \pi}\Big(H^2+\frac{k}{a^2} \Big)-\frac{1}{4\pi}\Big(\dot{H}-\frac{k}{a^2} \Big)\,,\label{FE2}
\end{eqnarray}
where $\rho_{\Lambda}=-P_{\Lambda}=\frac{\Lambda}{8\pi}$ \cite{Dolan:2013ft}. We can see that the energy density and pressure of the perfect fluid in the FRW universe spatially homogeneous: $i.e.$, $\tilde{\rho}=\tilde{\rho}(t)$, $\tilde{P}=\tilde{P}(t)$.
Regarding the apparent horizon (\ref{AH}) and with (\ref{dAH}), one can express the energy density (\ref{FE1}) and the pressure (\ref{FE2}) in terms of the apparent horizon radius and its time derivative
\begin{eqnarray}\label{P}
\tilde{\rho}=\frac{3}{8\pi R_{A}^{2}} \,,~~~~~~~~
\tilde{P}=-\frac{3}{8\pi R_{A}^{2}}+\frac{\dot{R}_{A}}{4\pi H R_{A}^{3}}\,.
\end{eqnarray}

In a FRW universe, the Misner-Sharp energy in Einstein gravity is obtained \cite{Cai:2005ra,Cai:2006rs,Gong:2007md,Hu:2015xva}, and given by
\begin{equation} \label{MSE}
E=\frac{R_A}{2}\,.
\end{equation}
From the definition of work density \cite{Hayward:1993wb,Hayward:1997jp}, we obtain
\begin{equation}\label{WPsi}
W:=-\frac{1}{2}h^{i j}T_{i j}=\frac{1}{2}(\tilde{\rho}-\tilde{P})=\frac{3}{8\pi R^2_A}-\frac{\dot{R}_A}{8\pi HR^3_A}\,,
\end{equation}
where $T_{i j}$ is the projection of the energy-momentum tensor $T_{\mu \nu}$ in the $(t,r)$ directions.
Using (\ref{MSE}) with (\ref{T1}) and (\ref{WPsi}), we get the first law of thermodynamics of the FRW universe
\begin{equation}\label{UFL}
dE=-TdS+W dV\,,
\end{equation}
where $S\equiv{A}/{4}=\pi R_A^2$ is the Bekenstein-Hawking entropy and $V=4\pi R_A^3/3$ is the thermodynamic volume of a $3$-dimensional sphere with radius $R_A$.
Furthermore, it is natural to extract the definition from the first law (\ref{UFL}), i.e.
\begin{equation}\label{WP}
 P:=W\,,
\end{equation}
so that, with $U := -E$, the first law can be rewritten as the usual form
\begin{equation}\label{StandFL}
dU=TdS-P dV\,.
\end{equation}
From this equation, we can see that the volume $V=4\pi R_A^3/3$ is also the thermodynamic volume, $i.e.$, conjugate to the thermodynamic pressure. From (\ref{T1}) to eliminate $\dot{R}_{A}$, and then substituting into (\ref{WPsi}) one can obtain the thermodynamic equation of state\footnote{The form of this equation of state is similar to the results obtained in our previous work \cite{Abdusattar:2021wfv,Kong:2021dqd,Abdusattar:2022bpg}. However, the thermodynamic pressure of the system here includes both the work density of the perfect fluid and the pressure defined by the cosmological constant. Notably, this system also exhibits no $P$-$V$ phase transition.}
\begin{eqnarray}\label{PVT}
P=\frac{T}{2R_{A}}+\frac{1}{8 \pi R_{A}^2}\,,~~~~~~~R_{A}=\left(\frac{3V}{4\pi}\right)^{1/3} \,.
\end{eqnarray}


\section{Thermodynamic Heat Engine Cycle Analysis in the FRW Universe}\label{sIII}

Thermodynamics, a pivotal branch of physics, deals with the conversion of energy and work within a system. Within this domain, heat is recognized as an energy form capable of being transferred from one body to another, or from one part of a body to another. Notably, a heat engine stands out as a physical system that transforms heat into mechanical energy through work, absorbing heat from a hotter region and releasing it to a colder one.

In this section, we adopt a novel perspective by treating the FRW universe as a heat engine and delving into its thermodynamic efficiency. To quantify the work performed by this cosmic heat engine, we utilize the $P$-$V$ diagram, which illustrates a closed path for the heat engine. To comprehensively study the characteristics of the FRW universe as a heat engine, we have meticulously selected two types of cycles: the Carnot cycle and the rectangular cycle. These two cycles serve as essential tools for gaining insights into the energy conversion and efficiency performance of the FRW universe when viewed as a heat engine. Through this innovative analytical framework, we aim to uncover the unique thermodynamic properties of the FRW universe and explore its potential mechanisms for energy conversion and work production.

It is worth clarifying that the thermodynamic cycles considered here should be interpreted as effective paths in the space of thermodynamic variables, rather than as literal cyclic evolutions of the cosmological spacetime. The heat-engine description relies on the assumption that, along these paths, thermodynamic quantities such as pressure, volume, and temperature remain operationally well defined. In a fully dynamical FRW background, this interpretation is therefore expected to apply within regimes where the evolution is sufficiently slow for a quasi-equilibrium description to hold.

\subsection{Carnot Cycle Analysis}

\begin{figure}[h]
 \begin{minipage}[t]{1\linewidth}
\includegraphics[width=6.5cm,height=6cm]{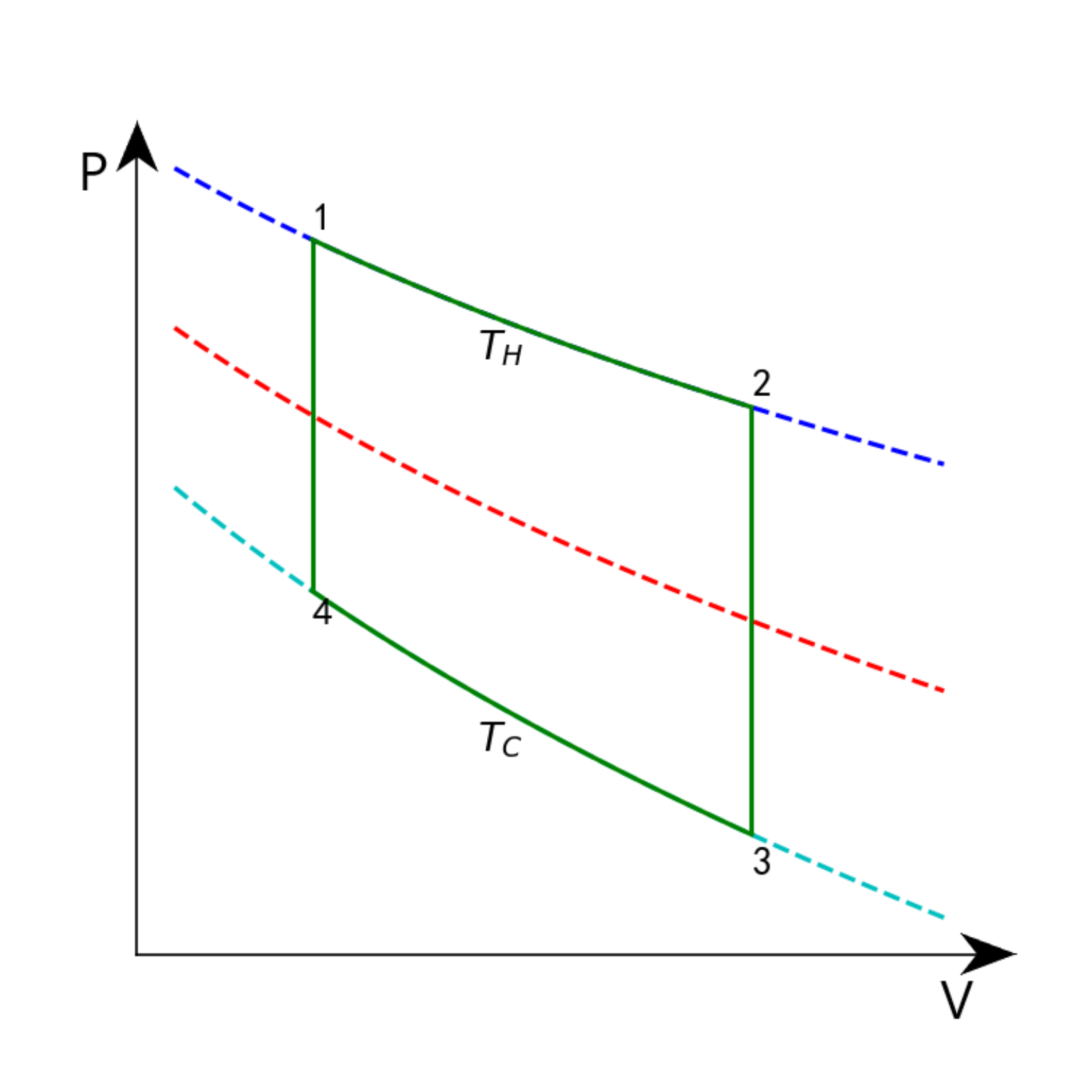}
 \end{minipage}
\caption{\footnotesize
{\bf $P$-$V$ diagram of the thermodynamic cycle for a Carnot heat engine in the context of the FRW universe:} Paths $1\rightarrow2$ and $3\rightarrow4$ are isothermal. Paths $2\rightarrow3$ and $4\rightarrow1$ are isochoric or adiabatic.}
\label{cyclePV1}
\end{figure}

In this subsection, we focus on the analysis of the Carnot heat engine cycle within the context of the FRW universe. The corresponding Carnot cycle diagram for the heat engine in the FRW universe is depicted in Fig.\ref{cyclePV1}. This diagram provides a visual representation of the pressure-volume changes that occur during the various stages of the cycle, enabling us to gain a deeper understanding of the energy conversion processes within the FRW universe when modeled as a heat engine.

$T_{H}$ and $T_{C}$ represent temperatures of the hot and cold regions respectively with $T_{H}>T_{C}$. The heats are absorbed from the hot region (state $1$ to state $2$) and released to the cold region (state $3$ to state $4$), which are denoted as $Q_H$ and $Q_C$ respectively. A total mechanical work $W$, which is produced by the heat engine, is given by $W=Q_H-Q_C$. Thus, the efficiency of the heat engine is defined by the ratio of total work and the amount of heat absorbed from the hot region, i.e. $\eta=\frac{W}{Q_{H}}$.
Using Eq.(\ref{PVT}) with the entropy $S=\pi R_A^2$, the heat absorbed from the hot region is
\begin{eqnarray}\label{CQH}
Q_H=\int^{S_{2}}_{S_{1}}T_H d S={\pi}T_H(R_{A2}^2-R_{A1}^2)\,.
\end{eqnarray}
Similarly, the heat released to the cold region is
\begin{eqnarray}\label{CQC}
Q_C=-\int^{S_{4}}_{S_{3}}T_C d S={\pi}T_C(R_{A3}^2-R_{A4}^2)\,.
\end{eqnarray}
In this case, the two adiabatic lines are just equal volume (or $R_A$) lines, so $R_{A1}=R_{A4}, R_{A2}=R_{A3}$,
and the total work is
\begin{equation}\label{Wm}
 W=Q_H-Q_C={\pi}(T_H-T_C)(R_{A2}^2-R_{A1}^2)\,.
\end{equation}

Thus the maximum efficiency of the Carnot engine becomes
\begin{eqnarray}\label{etaC}
\eta_C=\frac{W}{Q_H}=1-\frac{T_C}{T_H}\,,
\end{eqnarray}
which depends only on the two temperatures, as expected. One infers from the result (\ref{etaC}) that $0<\eta_C<1$ because $T_H>T_C$.

We also rewrite the efficiency of this Carnot heat engine in the following form
\begin{eqnarray}\label{etaC24}
\eta_C=1-\frac{T_4(P_4,R_{A1})}{T_2(P_1,R_{A2})}=1-\frac{R_{A2}(1-8\pi R_{A1}^2 P_4)}{R_{A1}(1-8\pi R_{A2}^2 P_1)}\,,
\end{eqnarray}
where we consider that the higher temperature $T_H$ and the lower temperature $T_C$ in our cycle correspond to $T_2$ and $T_4$, respectively.

\subsection{The Rectangular Cycle}

\begin{figure}[h]
 \begin{minipage}[t]{1\linewidth}
\includegraphics[width=6.5cm,height=6cm]{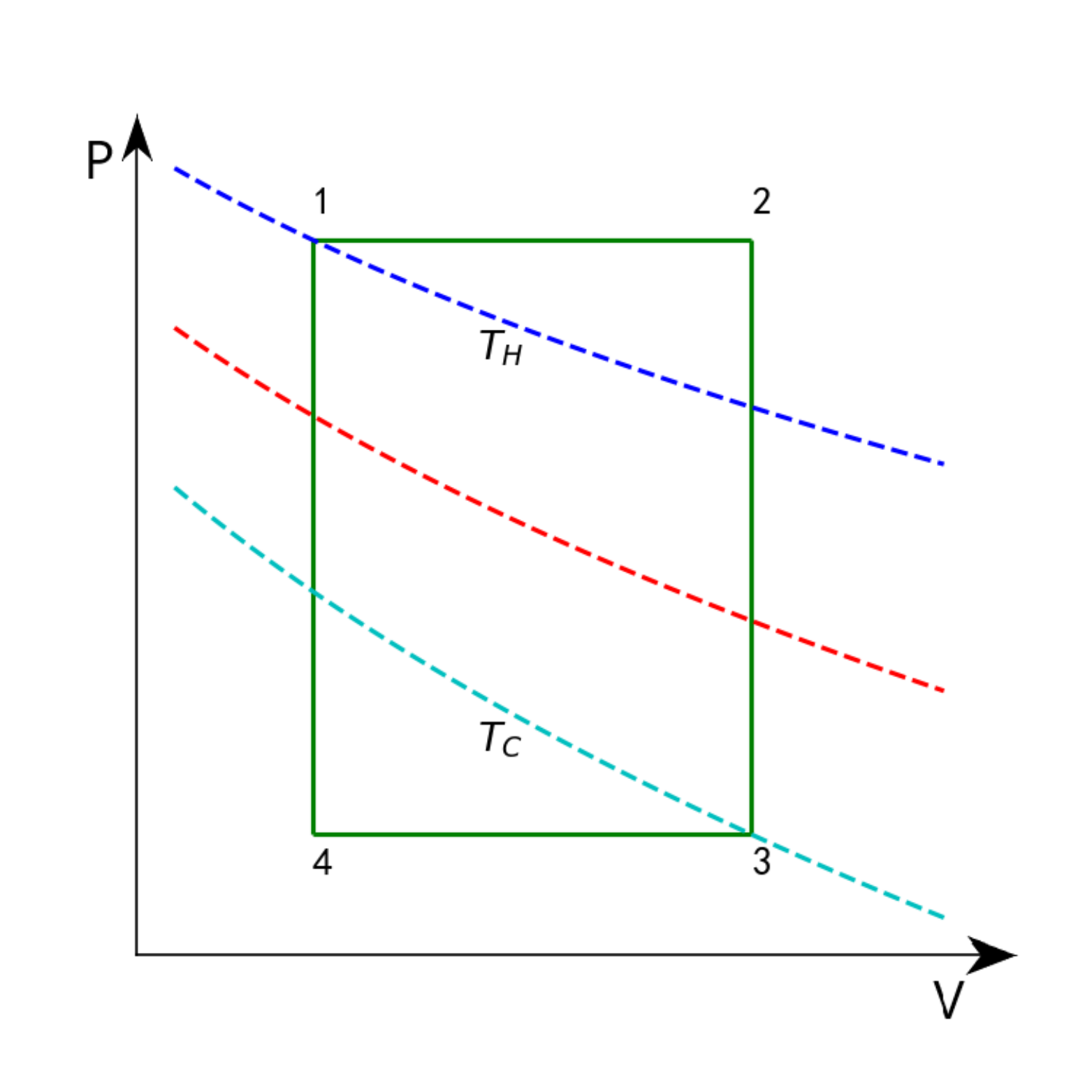}
 \end{minipage}
\caption{\footnotesize
{\bf $P$-$V$ diagram of the thermodynamic cycle for a rectangle heat engine in the context of the FRW universe:} Paths $1\rightarrow2$ and $3\rightarrow4$ are isobaric. Paths $2\rightarrow3$ and $4\rightarrow1$ are isochoric or adiabatic.}
\label{cyclePV2}
\end{figure}

In this subsection, we turn our attention to the analysis of the Rectangular heat engine cycle within the framework of the FRW universe. Inspired by Johnson's innovative work \cite{Johnson:2014yja} in devising a new cycle that incorporates two isobars and two isochores, where heat transfer occurs along the top and bottom lines, we now extend this concept to the FRW universe. In doing so, we introduce a novel heat engine cycle for the FRW universe, whose diagram is depicted in Fig.\ref{cyclePV2}. This new cycle offers a fresh perspective on energy conversion and efficiency within the cosmic context of the FRW universe.

The work done around this cycle is
\begin{equation}\label{NewW}
W=\triangle P_{4\rightarrow 1}~\triangle V_{1\rightarrow
2}=(P_1-P_4)(V_2-V_1)\,,
\end{equation}
where $P_{1}$ and $P_{4}$ denote the pressures at stages $1$ and $4$ respectively.

By combining the entropy $S=\pi R_A^2$ and (\ref{PVT}), we obtain the specific heat capacity of the FRW universe \cite{Abdusattar:2021wfv}
\begin{eqnarray}\label{CP}
C_{P}&=&\left(\frac{\partial {\cal H}}{\partial T}\right)_{P}=T\left(\frac{\partial S}{\partial T}\right)_{P} \\
&=&\frac{2{\pi}R_{A}^{2}(-1+8{\pi}R_{A}^2 P)}{1+8{\pi} R_{A}^2 P}=\frac{4{\pi}^{2}R_{A}^{3} T}{1+2{\pi}T R_{A}}\,,\nonumber
\end{eqnarray}
where ${\cal H}$ is the enthalpy of the FRW universe given by
\begin{equation}\label{enthalpyH}
 {\cal H}=U+P V=-\frac{R_A}{2}+\frac{4\pi R_A^3}{3}P \,.
\end{equation}
The heat absorbed along the upper isobaric line is given by
\begin{eqnarray}\label{NewQH}
Q_H&=&{\cal H}_2-{\cal H}_1=\int^{T_2}_{T_1}C_P dT \nonumber\\
&=& P_1(V_{2}-V_{1})-\frac{1}{2}({R_{A2}-R_{A1}})\,.
\end{eqnarray}
With the Eqs.(\ref{NewW}) and (\ref{NewQH}), the efficiency of the new heat engine can be derived as
\begin{eqnarray}\label{Neta}
\eta=\frac{W}{Q_H}&=&\frac{2(V_2-V_1)(P_1-P_4)}{2P_1 (V_{2}-V_{1})-({R_{A2}-R_{A1}})}\nonumber \\
&=& \left(1-\frac{P_4}{P_1}\right) \times \frac{1}{-\frac{R_{A2}-R_{A1}}{2P_1(V_2-V_1)}+1}\,. \,\,\,\,\,\,\,\,\,
\end{eqnarray}
We see that the efficiency $\eta$ of the new engine is always positive if $P_1>P_4$, $R_{A2}>R_{A1}$ (or $V_{2}>V_{1}$) and $P_1>\frac{R_{A2}-R_{A1}}{2(V_2-V_1)}$.

One can also compare the heat engine efficiency $\eta$ of rectangle cycle with the well-known Carnot efficiency $\eta_C$ and investigate the ratio ${\eta}/{\eta_C}$.
In the large apparent horizon limit, from (\ref{Neta}) and (\ref{etaC24}), we can obtain
\begin{equation}\label{detaR}
  \lim\limits_{R_{A2} \to \infty}\eta=\lim\limits_{R_{A2} \to \infty}\frac{\eta}{\eta_C}=1-\frac{P_4}{P_1}\,.
\end{equation}

\begin{figure}[h]
\subfigure[~The efficiency of the heat engine versus the pressure $P_1$ for FRW universe]{
\begin{minipage}[t]{6.5cm}
\includegraphics[width=6.5cm,height=5cm]{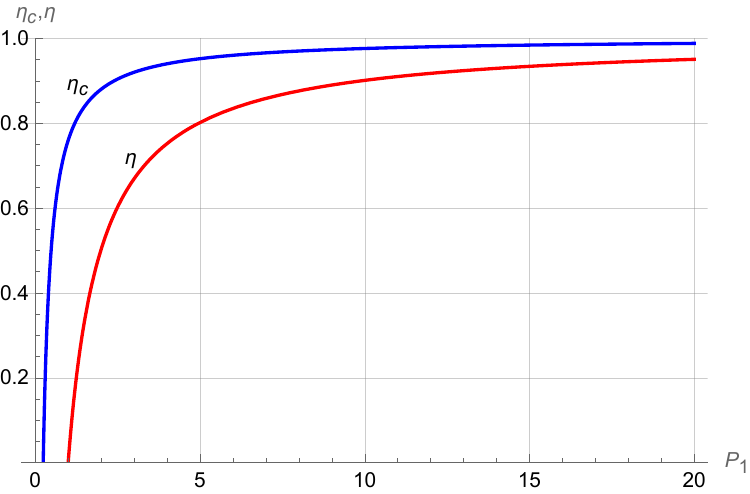}
\end{minipage}}
\subfigure[~The ratio ${\eta}/{\eta_C}$ versus the pressure $P_1$]{
\begin{minipage}[t]{6.5cm}
\includegraphics[width=6.5cm,height=5cm]{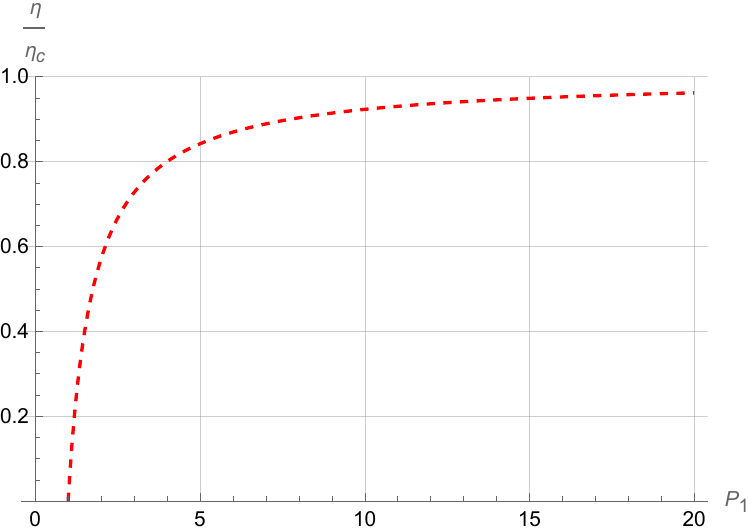}
\end{minipage}}
\caption{\small{Specific example with the parameters chosen as $P_4=1$, $R_{A1}=1$, $R_{A2}=4$.}}
\label{detaRP}
\end{figure}

To have a better understanding of the heat engine efficiency, we plotted the efficiency under the change of pressure $P_1$ as shown in Fig.\ref{detaRP}.

We can see that both the efficiency $\eta_C$ and $\eta$ will monotonously increase with the growth of pressure $P_1$, and approach the maximum efficiency, i.e. allowed by thermodynamics laws. The ratio of efficiencies ${\eta}/{\eta_C}$ also increases with the growth of pressure $P_1$. At the high pressure limit, we have
\begin{equation}\label{detaP}
  \lim\limits_{P_{1} \to \infty}\eta=\lim\limits_{P_{1} \to \infty}\frac{\eta}{\eta_C}=1\,,
\end{equation}
which implies that the efficiency of new heat engine cycle will approach the efficiency of Carnot cycle.

\section{Conclusions and Discussion}\label{sIV}

In this paper, we have investigated the thermodynamic properties of the Friedmann-Robertson-Walker (FRW) universe as a thermodynamic system. We treat the expansion of the FRW universe as a thermodynamic heat engine and have discussed its efficiency for Carnot cycle and rectangle cycle.
We have calculated the work done by the heat engine for both cycle, and have obtained the corresponding efficiencies. Besides, we have plotted the efficiency diagram of heat engine for FRW universe, and found that both the efficiency $\eta_C$ and $\eta$ will monotonously increase with the growth of pressure $P_1$, and approach the maximum efficiency, i.e. allowed by thermodynamics laws $0<\eta<\eta_C<1$. Moreover, we can see that the ratio of efficiencies holds ${\eta}/{\eta_C}<1$ and in the high pressure limit will infinitely approach $1$, so it coincides with the thermodynamical second law.

This research, by examining heat engines within the framework of FRW cosmology thermodynamics, has not only provided fresh perspectives on the evolution of cosmic temperature and the universe's overall history but also marked a pivotal advancement in our pursuit to comprehend the intricacies of cosmic evolution. By integrating thermodynamics with cosmology, our work holds the potential to yield invaluable insights that could catalyze further investigations in this domain. As we venture further into the unknown realms of the universe, we foresee this research serving as a cornerstone for pioneering discoveries and a more profound grasp of the fundamental forces and mechanisms that govern our universe.

Finally, it is worth emphasizing that the thermodynamic and heat-engine interpretation presented in this work should be understood as an effective description, whose physical applicability depends on the dynamical regime of the FRW universe. The efficiencies, cycles, and bounds discussed here are expected to be meaningful within phases where a quasi-equilibrium thermodynamic description associated with the apparent horizon remains operational. Explicitly acknowledging this domain of validity helps frame the results in a consistent physical context and clarifies the scope within which the heat-engine picture can be reliably applied.

Beyond the quasi-equilibrium regime implicitly assumed throughout this analysis, departures from the expected thermodynamic relations should be interpreted not as violations of thermodynamic principles, but as indications of the breakdown of the effective thermodynamic description itself. In such situations, quantities such as temperature, entropy, pressure, volume, or even thermodynamic cycles and efficiencies may remain formally definable while progressively losing a clear operational meaning associated with physical heat exchange and work. From this perspective, the limits of applicability of the heat-engine picture correspond to physically meaningful transitions in the dynamical behavior of the FRW universe, rather than inconsistencies of the framework. This highlights the importance of viewing the present construction as an effective description with a well-defined domain of validity.


\section{Acknowledgment}

The authors thank Profs. Y.~P.~Hu, Hongsheng Zhang, and Dr. S.~B.~Kong for their insightful discussions.
This work is supported by the National Natural Science Foundation of China (NSFC) under grant No. 12465012, the Kashi University high-level talent research start-up fund project under grant No. 022024002, and the Tianchi Talented Young Doctors Program of the Xinjiang Uyghur Autonomous Region.


\end{document}